\documentclass[preprint,amsmath,amssymb,superscriptaddress,aps,prfluids,longbibliography]{revtex4-2}
\usepackage{geometry}
\usepackage{comment}
\usepackage{enumerate}
\usepackage{amsmath,nccmath,stackengine,scalerel,calrsfs}
\usepackage{xcolor}
\usepackage{mathrsfs}
\usepackage{graphicx}
\usepackage{dcolumn}
\usepackage{float}
\usepackage{bm}

\usepackage[mathlines]{lineno}
\usepackage{xcolor}
\usepackage{color}
\usepackage[colorlinks=true,linktoc=page,citecolor=blue,linkcolor=blue,]{hyperref}
\DeclareMathAlphabet{\pazocal}{OMS}{zplm}{m}{n}
\stackMath
\newcommand\reallywidehat[1]{%
\savestack{\tmpbox}{\stretchto{%
  \scaleto{
     \scalerel*[\widthof{\ensuremath{#1}}]{\kern-.6pt\bigwedge\kern-.6pt}%
    {\rule[-\textheight/2]{1ex}{\textheight}}
  }{\textheight}%
}{0.5ex}}%
\stackon[1pt]{#1}{\tmpbox}%
}

\newcommand{\bB}{\textit{\textbf{B}}}

\newcommand{\bJ}{\textit{\textbf{J}}}

\newcommand{\bb}{\textit{\textbf{b}}}

\newcommand{\bj}{\textit{\textbf{j}}}
\newcommand{\bk}{\textit{\textbf{k}}}
\newcommand{\bp}{\textit{\textbf{p}}}
\newcommand{\bq}{\textit{\textbf{q}}}
\newcommand{\bu}{\textit{\textbf{u}}}
\newcommand{\bx}{\textit{\textbf{x}}}
\newcommand{\by}{\textit{\textbf{y}}}
\newcommand{\bz}{\textit{\textbf{z}}}

\newcommand{\bnabla}{{\bm \nabla}}

\begin{document}
\title{On the contribution of the Hall term in small-scale magnetohydrodynamic dynamo}
\author{Arijit Halder}
\author{Supratik Banerjee}
\email{sbanerjee@iitk.ac.in}
\affiliation{Department of Physics, Indian Institute of Technology, Kanpur 208016, India}	
\author{Anando G. Chatterjee}
\affiliation{Computer Centre, Indian Institute of Technology, Kanpur 208016, India}
\author{Manohar K. Sharma}
\affiliation{Department of Physics, Indian Institute of Technology, Kanpur 208016, India}	
		
\begin{abstract}
A detailed study of small-scale Hall magnetohydrodynamic dynamo has been performed both analytically and numerically. Assuming the magnetic field and the current to be separate fields, the contribution of the Hall term has been decomposed into two parts and their individual contributions have been studied separately. Calculating the scale-separated transfer rates described in Dar \textit{et. al.} (Physica D, 157 (207), 2001), it is found that the small-scale current fields are the primary contributors in sustaining large scale magnetic fields. Furthermore, the nature of the scale-to-scale fluxes are found to be globally intact with the ion inertial scale.

\end{abstract}

\maketitle
\section{Introduction}
A turbulent flow is characterized by its prevalent nonlinearity over viscous properties \citep{Frisch_1995}. For the Navier-Stokes equations, this nonlinearity is represented by the velocity advection term $\left(\bu\cdot\bnabla\right)\bu$ where $\bu$ is the fluid velocity \citep{Pope_2001,Tennekes_1972}. For a magnetohydrodynamic (MHD) fluid, the nonlinearity in momentum equation is represented by both $\left(\bu\cdot\bnabla\right)\bu$ and the Lorentz force term $\bJ \times \bB $ where $\bB$ is the magnetic field and $\bJ= \mu_0^{-1} \left(\bnabla \times \bB \right) $ is the current density \citep{Biskamp_2003, Galtier_2016}. In case of ordinary MHD turbulence, the induction equation takes the form 
\begin{equation}
\frac{\partial \bB}{\partial t} = \bnabla \times \left(\bu \times \bB\right)+\eta\nabla^2 \bB,
\end{equation} 
where the first term on the right hand side represents the nonlinear contribution and the second term represents the diffusion, $\eta$ being the magnetic diffusivity. Similar to hydrodynamic turbulence, in MHD we also observe cascade of total energy (kinetic + magnetic) which is an inviscid invariant of the system. It is important to note that, in the ordinary MHD model ions and electrons move with the same velocity (the fluid velocity) \citep{Bittencourt_2013, Galtier_2016}. However, this does not remain valid for length scales close or smaller to the ion inertial length $d_i$. For those length scales, it is important to include the effect of finite difference between the ion and electron velocities \citep{Belmont_2013}. This effect is incorporated in the induction equation by adding a nonlinear term $- \bnabla \times \left(\bJ \times \bB\right)/ne$, often called the Hall term, where $n$ and $e$ are the number density of the electrons or ions and the electronic charge respectively and the corresponding model is known as Hall MHD (HMHD). In terms of $\bb$ ($= \bB/\sqrt{\mu_0 \rho}$ with $\rho$ being the mass density of MHD fluid), the magnetic field normalized to a velocity and $\bj= \bnabla \times \bb$, the induction equation simply gives the evolution of $\bb$ and corresponding Hall term can be written as $- d_i\bnabla \times\left(\bj \times \bb\right)$. Despite having different ion and electron velocities, HMHD is a monofluid model and cannot capture all the essential aspects of sub-ion scale plasma dynamics including the kinetic effects. However, HMHD turbulence is often found quite useful to probe into the sub-ion scale turbulence in the space and astrophysical plasmas \citep{Andres_2018, Papini_2019}. 

Several aspects of HMHD turbulence (both incompressible and compressible) have been explored analytically, numerically and also using \textit{in-situ} spacecraft data  during the last two decades. Besides systematic studies covering the basic features like energy cascade, power-spectrum, universal scaling, dissipative anomaly etc. \citep{Galtier_2008, Banerjee_2016a, Meyrand_2012, Galtier_2018, Galtier_2006, Ferrand_2019}, some interesting studies have also been made exploring the importance of HMHD turbulence in the heating of the solar wind and other space plasmas \citep{Galtier_2007, Hellinger_2018, Andres_2019, Bandyopadhyay_2020}. Dynamo mechanism is essential to explain long lasting magnetic fields in astrophysical plasmas \citep{Choudhuri_1998, Brandenburg_2012, Rincon_2019}. Whereas a sustained large-scale magnetic field can be explained in terms of mean-field MHD and HMHD, a turbulent dynamo mechanism is often useful to explain the development and energization of magnetic field fluctuations of length scales inferior to the external forcing scale (small-scale dynamo) \citep{Mininni_2003,Mininni_2005a,Mininni_2005b,Mininni_2007,Lingam_2016,Kumar_2013}. The role of Hall effect in small-scale dynamo has been studied using numerical simulations both for helical and non-helical forcings \citep{Mininni_2007, Gomez_2010}. Using a resolution of $256^3$, both studies reported a backscatter (to larger scales) of magnetic energy due to the Hall term. The flux rates due to the Hall term were found to be approximately 10-15 times weaker than the MHD flux rates. The Hall contribution was calculated from the entire $- d_i\bnabla \times\left(\bj \times \bb\right)$ term and the corresponding flux rates were simply interpreted as transfers between $\bb$-fields of different Fourier modes using $\bj= \bnabla \times \bb$. However, recent studies showed that a clear phenomenological picture can be obtained if we decompose the $\bnabla\times(\bu\times\bb)$ term in ordinary MHD as 
$-\left(\bu\cdot\bnabla\right)\bb$ and $\left(\bb\cdot\bnabla\right)\bu$ \citep{Verma_2004, Plunian_2019}. For the sake of uniformity with MHD nonlinearities and to get further physical insight, it is thus reasonable to decompose the Hall contribution in the induction equation in a similar way. Such a decomposition results into two contributions $-d_i \bb\cdot\left(\bb\cdot\bnabla\right)\bj$ and $d_i \bb\cdot\left(\bj\cdot\bnabla\right)\bb$ in the magnetic energy evolution equation ($\bb\cdot\partial_t\bb$) representing $\bj$-to-$\bb$ and $\bb$-to-$\bb$ transfer respectively. 

In this paper, using 3D direct numerical simulation (DNS) of $256^3$ spatial resolution, we present a systematic study comparing the individual contribution of the aforementioned sub-parts of the Hall term in the small scale dynamo action. For the sake of computational convenience, here we have calculated the fluxes using mode-to-mode (M2M) transfers \citep{DAR_2001, Verma_2004}. Such a methodology is particularly useful in the study of dynamo as it helps in classifying different types of energy transfers depending on the length scales (or equivalently the wavevectors). In order to recognise the Hall contribution unambiguously, we have employed a $\nabla^4$- hyperviscosity. Similar to the previous studies, a clear signature of backscatter of magnetic energy due to the Hall term is also observed. Of all possible contributions, the transfer from small scale $\bj$-fields to large scale \footnote{The large scale field here signifies the magnetic field between the forcing and the seed field scale} $\bb$-fields is found to be the most efficient one. 
Furthermore, the maximum of the total transfer rate to large scale $\bb$-field shifts towards larger scales as we go on increasing $d_i$.

The rest of the paper is organised as follows: In Sec.~\ref{sec:Governing equations}, we describe the basic equations of incompressible HMHD and expressions for spectral fluxes using M2M formalism. 
Sec.~\ref{methodology} contains the description of the code and details of the numerical simulations. In Sec.~\ref{fluxes} and \ref{dependence}, we present different contributions to magnetic energy transfer to large scale $\bb$-fields and their variation with $d_i$ respectively. Finally in Sec.~\ref{Conclusion}, we discuss our findings and conclude.

\section{\bf{Mode-to-mode energy transfers and spectral fluxes}}
\label{sec:Governing equations}
We start with the incompressible Hall MHD equations \citep{Galtier_2016}
\begin{align}
\partial_t \bu -  \nu\nabla^2 {\bu} &= -(\bu\cdot \bnabla)\bu +  \left(\bb\cdot\bnabla\right)\bb  -\bnabla \left(p+\frac{b^2}{2}\right) + {\textbf{\textit f}} , \label{eq:u1}\\
\partial_t \bb - \eta\nabla^2 {\bb} &= -\left(\bu\cdot\bnabla\right)\bb+\left(\bb\cdot\bnabla\right)\bu +d_i\left(\bj\cdot\bnabla\right)\bb -d_i\left(\bb\cdot\bnabla\right)\bj,\label{eq:b1}\\
\bnabla \cdot \bu & = 0, \quad   \bnabla \cdot \bb =  0, \label{eq:div}
\end{align}
where $p$ is the kinetic pressure (normalized by the constant fluid density $\rho$), {\textbf{\textit f}} is the external force per unit mass, $\nu$ and $\eta$ denote the kinematic viscosity and the magnetic diffusivity respectively. As mentioned in the introduction, the last two terms on the \textit{r.h.s.} of (\ref{eq:b1}) correspond to the Hall effect. The total energy $E\ (\ =\int \left(u^2+b^2\right)/2\ d\tau)$ is an inviscid invariant of both incompressible MHD and HMHD \citep{Mininni_2003,Mininni_2005a,Servidio_2008}.

In order to investigate the detailed conservation of energy, one requires the evolution equations for modal energy densities $E^u_{\bk}\ (\ =\bu_\bk\cdot\bu_\bk^*/2)$ and $E^b_{\bk}\ (\ = \bb_\bk\cdot\bb_\bk^*/2)$. First, we calculate the Fourier transforms of \eqref{eq:u1} and \eqref{eq:b1} and then the corresponding complex conjugates. Finally, using straightforward algebra, for inertial zone, one can show 
\begin{align}
\partial_t  E^u_\bk  &= \frac{1}{2}  \int\limits_{\bp}  \int\limits_{\bq} 
\left[ S_E^{uu}\left(\bk|\bp,\bq\right)+S_E^{ub}\left(\bk|\bp,\bq\right) \right]  \delta\left( \bk + \bp + \bq \right) d {\bp} \ d {\bq},\label{Energy_Fourier_timeu2}\\
\partial_t  E^b_\bk  &= \frac{1}{2}  \int\limits_{\bp}  \int\limits_{\bq} 
\left[S_E^{bb}\left(\bk|\bp,\bq\right)+S_E^{bu}\left(\bk|\bp,\bq\right) + d_i S_E^{\pazocal{H}}\left(\bk|\bp,\bq\right) \right]  \delta \left( \bk + \bp + \bq \right) d {\bp} \ d {\bq},
\end{align} 
where $S_E^{xy}\left(\bk|\bp,\bq\right)$ is the generic form of combined transfer rate of energy from $\by_\bp$ and $\by_\bq$ to $\bx_\bk$ and can be mathematically expressed as 
\begin{equation}
|S_E^{xy}\left(\bk|\bp,\bq\right)| = \Im \left[\left(\bp\cdot\bz_\bq\right)\left(\by_\bp\cdot\bx_\bk\right)+\left(\bq\cdot\bz_\bp\right)\left(\by_\bq\cdot\bx_\bk\right) \right],\label{combined_generic}
\end{equation}
where $\bx$, $\by$ and $\bz$ can be $\bu$ and $\bb$. The algebraic sign becomes negative only when $\bx$ and $\by$ represent the same field. Note that, the $\bz$ fields do neither donate nor receive any energy in the process but only act as a mediator. $S_E^{\pazocal{H}}\left(\bk|\bp,\bq\right)$ is the combined magnetic energy transfer rate due to the Hall term only and can be mathematically expressed as  
\begin{equation}
S_E^{\pazocal{H}}\left(\bk|\bp,\bq\right)=\Im\left[\left(\bp\cdot\bb_\bq\right)\left(\bj_\bp\cdot\bb_\bk\right)+\left(\bq\cdot\bb_\bp\right)\left(\bj_\bq\cdot\bb_\bk\right)-\left(\bp\cdot\bj_\bq\right)\left(\bb_\bp\cdot\bb_\bk\right)-\left(\bq\cdot\bj_\bp\right)\left(\bb_\bq\cdot\bb_\bk\right)\right].\label{comb_Hall_k}
\end{equation}

For ordinary MHD as well as HMHD,  $S_E^{uu}\left(\bk|\bp,\bq\right)$ and $S_E^{bb}\left(\bk|\bp,\bq\right)$ individually satisfy triadic conservations whereas $S_E^{ub}\left(\bk|\bp,\bq\right)$ and $S_E^{bu}\left(\bk|\bp,\bq\right)$ together follow a conservation in each triad. Using \eqref{comb_Hall_k}, one can also obtain 
\begin{equation}
S_E^{\pazocal{H}}\left(\bk|\bp,\bq\right)+S_E^{\pazocal{H}}\left(\bp|\bq,\bk\right)+S_E^{\pazocal{H}}\left(\bq|\bk,\bp\right)=0,  \label{consH}
\end{equation}
where $S_E^{\pazocal{H}}\left(\bp|\bq,\bk\right)$ and $S_E^{\pazocal{H}}\left(\bq|\bk,\bp\right)$ are obtained under the cyclic permutation of the wavevectors. Thus the detailed conservation of total energy in HMHD can be written as $$\partial_t\left(E_\bk^u+E_\bk^b+E_\bp^u+E_\bp^b+E_\bq^u+E_\bq^b\right)=0.$$
According to Kraichnan, the triads constitute the fundamental unit of energy transfer in the Fourier space \citep{Kraichnan_1958, Kraichnan_1959}. However, some previous studies further decomposed the combined transfer rate as a sum of two M2M transfer rates as in the following manner: \begin{equation}
S^{xy}_{E}(\bk|\bp,\bq) = T_{E}(\by_\bp\overset{\bz_\bq}{\longrightarrow}\bx_\bk)+T_{E}(\by_\bq\overset{\bz_\bp}{\longrightarrow}\bx_\bk), \label{eq:SversusT}
\end{equation}
where $T_{E}(\by_\bp\overset{\bz_\bq}{\longrightarrow}\bx_\bk)$ is interpreted as the M2M transfer rate of energy from the giver field $\by_\bp$ to the receiver field $\bx_\bk$ with $\bz_\bq$ acting as a mediator \citep{Dar_2000, DAR_2001, Verma_2004, Plunian_2019}. Note that, the M2M transfer rates are always defined upto an additive arbitrary constant \citep{Dar_2000, Plunian_2019}. However, the calculation of flux and shell-to-shell transfer rates are not affected by such constants which cancel out in individual triads.
As mentioned before, in this work, the Hall term is decomposed in two parts and the combined energy transfer rate corresponding to the $\bk$-th mode is given by 
\begin{equation}
S_E^{\pazocal{H}}\left(\bk|\bp,\bq\right)=\Sigma_E^{bj}\left(\bk|\bp,\bq\right)+\Sigma_E^{bb}\left(\bk|\bp,\bq\right),\label{hallalt2}  
\end{equation} where
\begin{align}
\Sigma_E^{bj}\left(\bk|\bp,\bq\right)&=\Im\left[\left(\bp\cdot\bb_\bq\right)\left(\bj_\bp\cdot\bb_\bk\right) + \left(\bq\cdot\bb_\bp\right)\left(\bj_\bq\cdot\bb_\bk\right)\right]\;\;\;\;\text{and}\label{hallalt2bj}\\
\Sigma_E^{bb}\left(\bk|\bp,\bq\right)&= - \Im\left[\left(\bp\cdot\bj_\bq\right)\left(\bb_\bp\cdot\bb_\bk\right) + \left(\bq\cdot\bj_\bp\right)\left(\bb_\bq\cdot\bb_\bk\right)\right]\label{hallalt2bbk}
\end{align}
represent the $\bj$-to-$\bb$ and $\bb$-to-$\bb$ transfers respectively. In analogy with the ordinary MHD case, here one can immediately understand that $\Sigma_E^{bj}$ and $\Sigma_E^{bb}$ are originated from the terms $-\left(\bb\cdot\bnabla\right)\bj$ and $\left(\bj\cdot\bnabla\right)\bb$ respectively. From \eqref{hallalt2bbk}, one can immediately derive
\begin{equation}
\Sigma_E^{bb}\left(\bk|\bp,\bq\right)+\Sigma_E^{bb}\left(\bp|\bq,\bk\right)+\Sigma_E^{bb}\left(\bq|\bk,\bp\right)=0,\label{hallalt2bbconsv}
\end{equation}
and combining \eqref{consH} and \eqref{hallalt2bbconsv} one can finally obtain
\begin{equation}
\Sigma_E^{bj}\left(\bk|\bp,\bq\right)+\Sigma_E^{bj}\left(\bp|\bq,\bk\right)+\Sigma_E^{bj}\left(\bq|\bk,\bp\right)=0.\label{hallalt2bjconsv}
\end{equation}
\begin{figure}[H]
   \centering
    \includegraphics[width=0.5\textwidth]{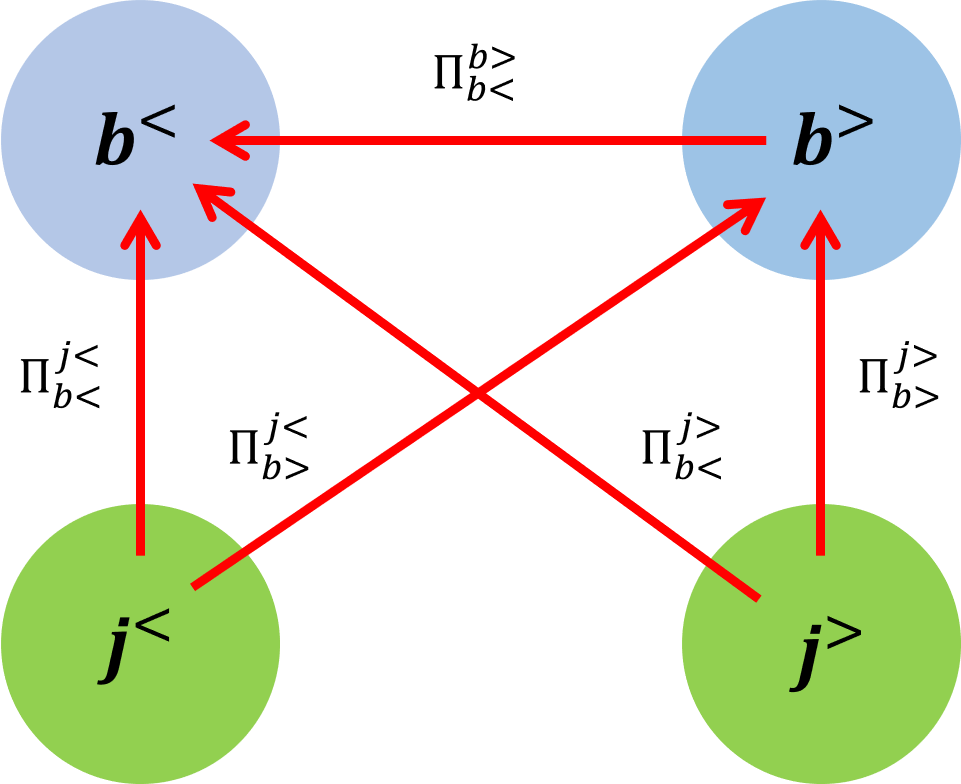}
    \caption{Schematic diagram of various magnetic energy fluxes due to the Hall term.}
    \label{fluxes_diagram}
\end{figure}
\noindent In this paper, we are interested in the contribution of the Hall term in the dynamo action. We are therefore only concerned with the feeding of large scale magnetic fields and hence we numerically calculate the fluxes $\Pi^{b>}_{b<}$, $\Pi^{j<}_{b<}$ and $\Pi^{j>}_{b<}$ and their mathematical expressions are given by
\begin{align}
    \Pi^{b>}_{b<}(k_0)&=-d_i\sum_{|\bk|\leq k_0}\sum_{|\bp|> k_0}\Im\left[\left(\bp\cdot\bj_{\bq}\right)\left(\bb_{\bp}\cdot\bb_{\bk}\right)\right],\label{flux_bibo_hall}\\
    \Pi^{j<}_{b<}(k_0)&=d_i\sum_{|\bk|\leq k_0}\sum_{|\bp|\leq k_0}\Im\left[\left(\bp\cdot\bb_{\bq}\right)\left(\bj_{\bp}\cdot\bb_{\bk}\right)\right],\label{flux_jibi_hall}\\
    \Pi^{j>}_{b<}(k_0)&=d_i\sum_{|\bk|\leq k_0}\sum_{|\bp|> k_0}\Im\left[\left(\bp\cdot\bb_{\bq}\right)\left(\bj_{\bp}\cdot\bb_{\bk}\right)\right],\label{flux_jobi_hall}
\end{align}
where $\Pi^{y<}_{x>}(k_0)$ is the net energy transfer rate from all the modes inside the sphere of radius $k_0$ of field $y$ to all modes outside $k_0$ of field $x$ (see Fig.~(\ref{fluxes_diagram})). Finally, the total flux to the large scale $\bb$-field can simply be calculated using 
\begin{equation}
    \Pi_{b<}=\Pi^{b>}_{b<}+\Pi^{j<}_{b<}+\Pi^{j>}_{b<}.
\end{equation}

\section{Numerical results}
\label{numerics}
\subsection{Methodology}
\label{methodology}
Here we perform a three-dimensional direct numerical simulation (DNS) of \eqref{eq:u1}-\eqref{eq:div} for $d_i=0.05$, $0.1$ and $0.2$. The simulations are performed using a parallel pseudo-spectral code based on the parallelization scheme by \citet{Mortensen_2016}. The box length is taken to be $2\pi$ and periodic boundary conditions have been assumed in all three spatial directions. In order to remove aliasing error, ${2}/{3}$-dealising rule has been used so that the maximum available wavenumber $k_{max}\approx 85$ \citep{Orszag_1971}. The equations are evolved using a 4-th order Runge-Kutta (RK4) method with a fixed time step $dt\ (\ =5\times 10^{-4})$. The flow is forced with a non-helical $(\bnabla\times{\textbf{\textit f}}\perp {\textbf{\textit f}})$ Taylor-Green (TG) forcing ${\textbf{\textit f}}= (sin(k_0x)cos(k_0y)cos(k_0z), -cos(k_0x)sin(k_0y)cos(k_0z), 0)$ with $k_0=2$. 
\begin{figure}[H]
    \hspace{0.8cm}
    \includegraphics[width=\linewidth]{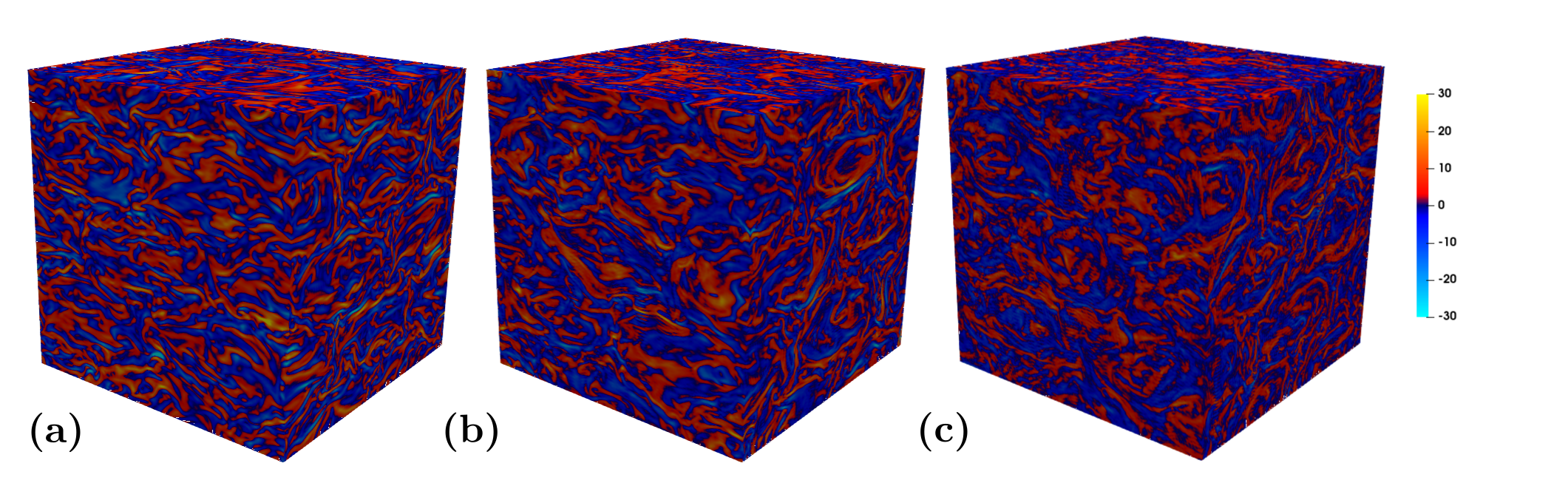}
    \caption{$x$-component of current density for (a) $d_i=0.05$, (b) $d_i=0.1$ and (c) $d_i=0.2$ at $t=45$.}
\end{figure}
For the sake of finding a clear Hall contribution, here we implement fourth order hyperdissipative operators for dissipation. Following general prescription of dynamo simulation, initially a pure hydrodynamic run (with hyperviscosity) was made to reach a turbulent steady state. After that, a random seed magnetic field was introduced at small scales $(k>10)$ such that the initial ratio of magnetic to kinetic energy was $\sim 10^{-3}$. Finally, the full HMHD equations were solved until the magnetic energy reaches saturation. Note that, the magnetic Prandtl number $(P_m)$ for our simulations is kept at unity. 
\begin{figure}[H]
    \centering
    \includegraphics[width=0.9\linewidth]{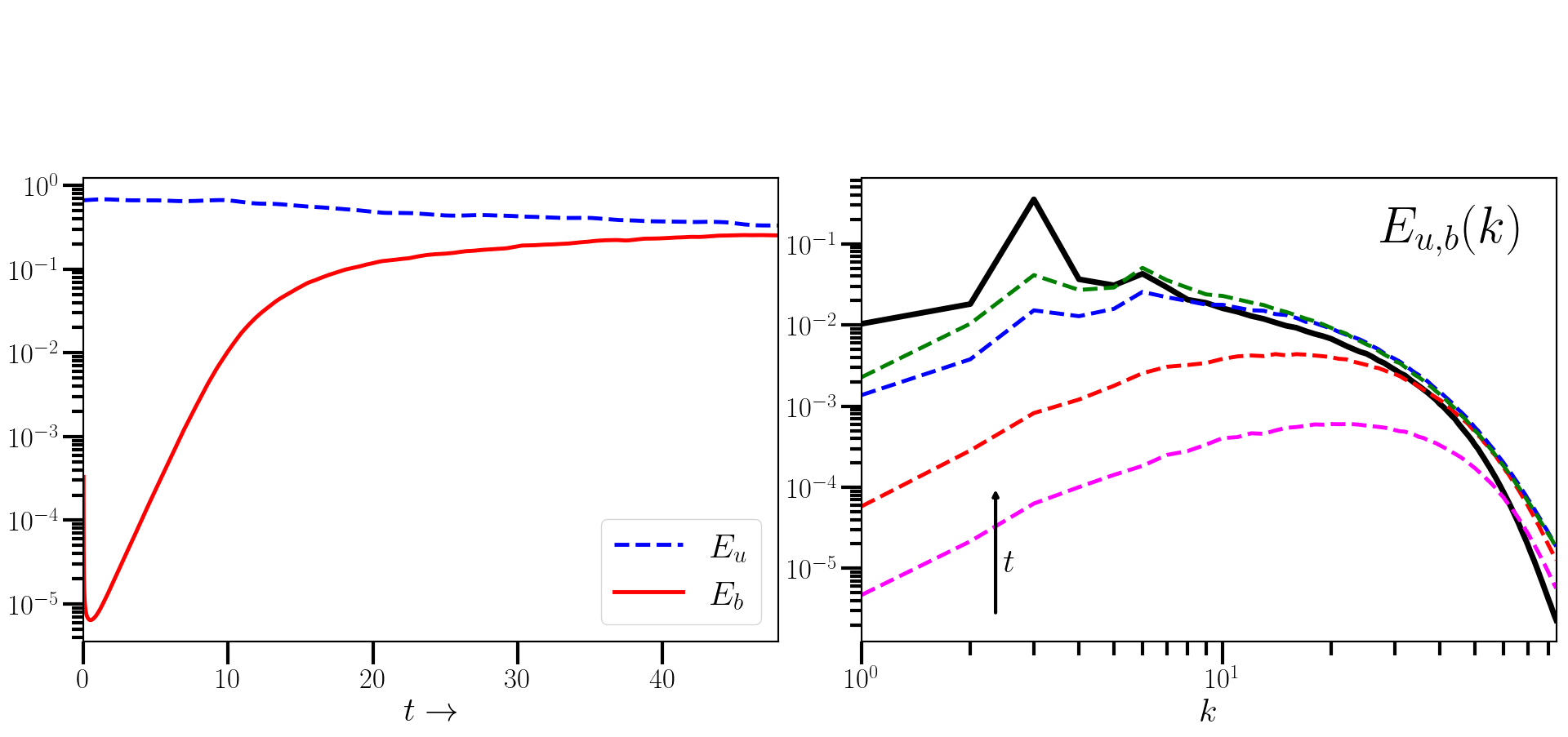}
    \caption{(left panel) Total kinetic energy (blue dashed curve) and total magnetic energy (red solid curve) vs. time. (right panel) Power spectra for magnetic energy for various times (dashed curves). The black solid curve corresponds to power spectra for kinetic energy at $t=45$.}
    \label{figeub}
\end{figure}
Left panel of Fig.~(\ref{figeub}) shows the time evolution of kinetic and magnetic energies whereas the right panel shows their spectral distributions. After an initial transient period, the magnetic energy grows exponentially. In this regime, the magnetic fields are approximately frozen in the bulk flow and hence the corresponding dynamo can be characterized as a kinematic-MHD dynamo. As time progresses, the electron flow starts to get decoupled from the bulk and another regime of kinematic-HMHD dynamo is reached where the magnetic field lines are effectively frozen in the electron fluid. Finally, the magnetic field saturates where it becomes strong enough to affect the motion of the flow thus leading to a non-kinematic dynamo. From the spectral distribution, it is easy to verify that large scale ($k<10$) magnetic fields grows with time. However, the large scale magnetic field grows much faster than small scale magnetic fields which effectively saturates near $t=20$.
Both of these observations are consistent with previous results on dynamo simulations in HMHD \cite{Mininni_2007, Gomez_2010}.

\subsection{Spectral fluxes}
\label{fluxes}
To get an estimate between the MHD and Hall transfers, we first plot the total flux rates. For our calculations, we have chosen the following radii: 2.00, 4.00, 8.00, 9.19, 10.56, 12.12, 13.93, 16.00, 18.40, 21.11, 24.25, 27.86, 32.00, 36.75, 42.22, 48.50, 55.71, 64.00 and 128.00. Such a logarithmic choice of radii is adopted in order to populate the inertial range where a power law scaling is expected to hold \citep{Kumar_2017}. 
\begin{figure}[H]
    \centering
    \includegraphics[width=\textwidth]{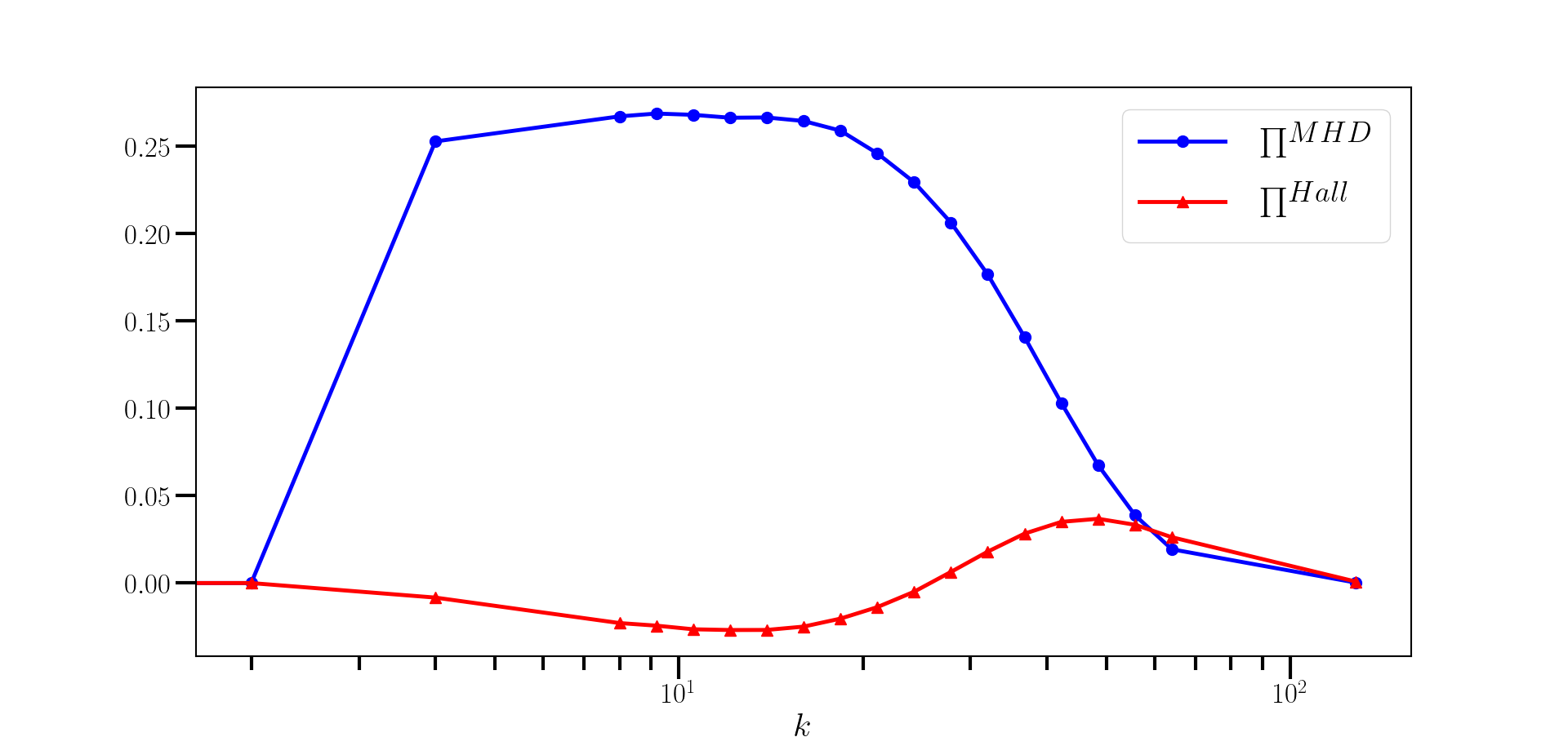}
    \caption{Total MHD vs. Hall flux rates for $d_i=0.05$.}
    \label{mhdvhmhd}
\end{figure}
\noindent Fig.~(\ref{mhdvhmhd}) shows the total transfer rates for magnetic energy, calculated using the M2M formalism described before. The Hall flux rate is found to be roughly 8 times weaker than the corresponding MHD flux rate which is clearly an improvement over earlier studies using ordinary fluid viscosities. As mentioned before, a clear backscatter of magnetic energy due to the Hall effect is observed for wavenumbers smaller than Hall wavenumber $(\approx 20)$. In order to further probe the aforesaid backscatter, we compute scale-classified fluxes using \eqref{flux_bibo_hall}-\eqref{flux_jobi_hall}.
\begin{figure}[H]
\hspace{-1.6cm}
    \includegraphics[width=1.1\linewidth, height=11.3cm, keepaspectratio=false]{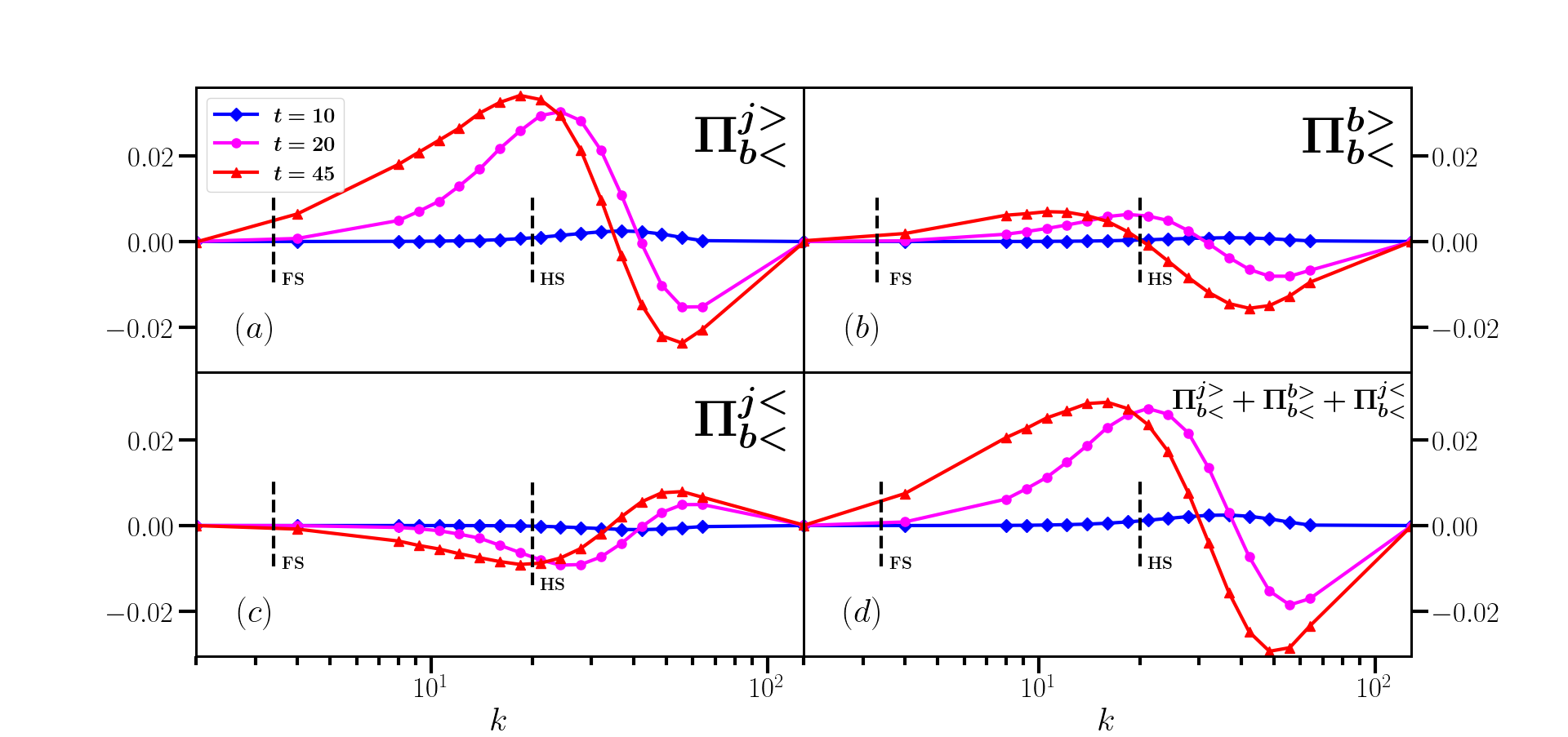}
    \caption{Various scale-to-scale fluxes important for dynamo at different times. FS = Forcing Scale, HS = Hall scale.}
    \label{flux_various}
\end{figure}
Fig.~(\ref{flux_various}) shows the flux rates corresponding to all possible channels responsible for the feeding of the large-scale magnetic field for three different times $t=10,\ 20\ \text{and}\ 45$. Such choices of time steps have been made in order to elucidate the role of Hall effect at kinematic-MHD dynamo regime, Hall saturated nonlinear dynamo and finally at the stage of magnetic energy saturation respectively. Fig.~(\ref{flux_various}a) shows the variation of $\Pi^{j>}_{b<}$ with $k$. A positive flux rate directly represents the net transfer of magnetic energy from small scale $\bj$-field to large scale $\bb$-field (dynamo action) and vice-versa. At $t=10$, the flux rate is slightly positive beyond the Hall scale. As time progresses, the transfer rate becomes larger. Interestingly, the flux rate is not sign definite and in fact it changes sign for length scales superior to the Hall scale. At $t=20$, the flux changes sign around $k=40$ whereas for $t=45$ the sign change occurs around $k=30$. Hence, with time, the transition wavenumber is found to shift towards the Hall wavenumber $(k=20)$. For $t=45$, where the magnetic energy saturates (see Fig.~(\ref{figeub})), the maximum of the positive flux rate peaks around the Hall scale. Fig.~(\ref{flux_various}b) shows the variation of $\Pi^{b>}_{b<}$ with wavenumbers. A positive flux rate therefore signifies a net transfer of magnetic energy from small to large scale $\bb$-field. Similar to $\Pi^{j>}_{b<}$, at $t=10$, the transfer rate is found to be slightly positive beyond the Hall scale.
At $t=20$, the flux rate increases and becomes negative beyond $k=40$. Note that, a negative flux rate in this case signifies a transfer towards small scale $\bb$-field \textit{i.e.}, a direct cascade. For $t=45$, the positive peak of the flux rate is found to shift further towards smaller wavenumbers while the change of sign occurs around Hall scale ($k=20$). Fig.~(\ref{flux_various}c) shows the variation of $\Pi^{j<}_{b<}$ with $k$. It denotes the transfer of magnetic energy from large scale $\bj$-field to large scale $\bb$-field. At $t=10$, in contrast to the previous cases, the flux rate is found to be slightly negative beyond the Hall scale. A negative flux rate signifies that the large scale magnetic field loses energy instead of gaining. Thus, in the kinematic regime, the large scale current fields remove energy from the large scale magnetic fields rather than feeding them. As time progresses, both positive and negative flux rates increase. At $t=20$, it changes sign from negative to positive around $k=45$ whereas at $t=45$ the sign change takes place around $k=30$. In addition, the peak of negative flux rate is also found to move towards the Hall scale. Of all the aforementioned transfer rates, clearly $\Pi^{j>}_{b<}$ is the largest. It signifies that the small scale $\bj$-fields are the primary contributor in sustaining large scale $\bb$-fields. Similarly, $\Pi^{b>}_{b<}$ also contributes in dynamo action through the transfer of magnetic energy from small scale $\bb$-fields to large scale $\bb$-fields. However, $\Pi^{j<}_{b<}$ does not directly seem to enhance the large scale magnetic fields, rather it facilitates the growth small scale magnetic fields. The sum of all aforesaid transfers is the net transfer rate of magnetic energy to the large scales. Therefore, a positive value denotes that a dynamo action can be maintained through the scale-to-scale transfer from various scales. Fig.~(\ref{flux_various}d) shows the variation of total flux rate $\Pi^{j>}_{b<}+\Pi^{b>}_{b<}+\Pi^{j<}_{b<}$ with $k$. 
Interestingly, it shows similar behaviour to that of $\Pi^{j>}_{b<}$. This is expected since out of all three individual fluxes, $\Pi^{j>}_{b<}$ is the most dominating one. 
\subsection{Dependence of transfer rates on Hall scale}
\label{dependence}
\begin{figure}[H]
\hspace{-2.7cm}    
\includegraphics[width=1.2\textwidth]{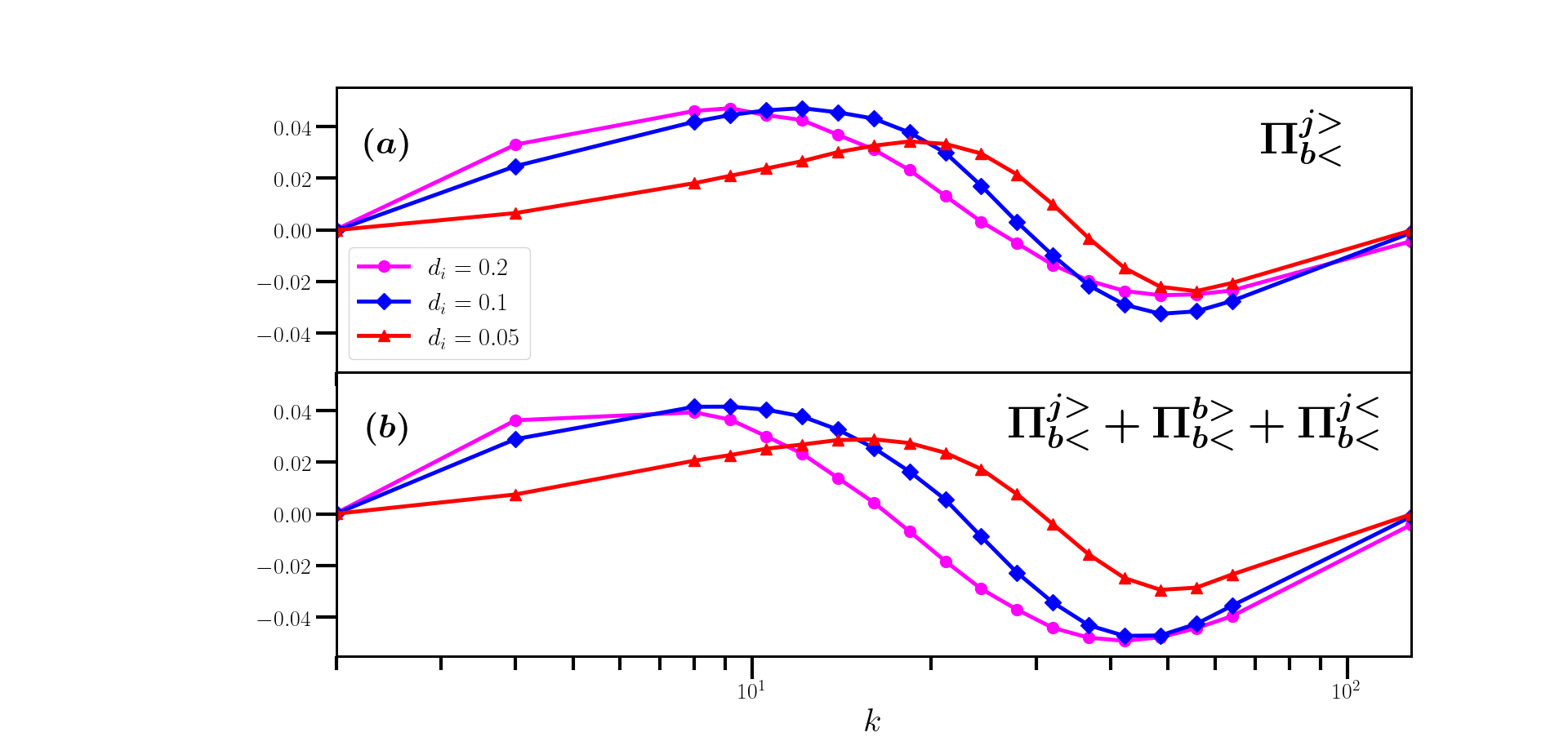}
    \caption{(a) $\Pi^{j>}_{b<}$ and (b) $\Pi^{j>}_{b<}+\Pi^{b>}_{b<}+\Pi^{j<}_{b<}$ vs Hall scale ($d_i$) at $t=45$.}
    \label{jobi-vs-total-di}
\end{figure}

In this section, we study the variation of aforementioned flux rates with the Hall parameter ($d_i$). We have performed three simulations with Hall parameters $d_i=0.05,\ 0.1\ \text{and}\ 0.2$ with the corresponding Hall wavenumbers given by 20, 10 and 5 respectively. Note that, for our simulations, the forcing wavenumber $k_f\approx 3.4$.
Instead of studying the variation of fluxes with $d_i$ for different times, here we only consider their values at magnetic energy saturation ($t=45$). Fig.~(\ref{jobi-vs-total-di}a) shows the variation of $\Pi^{j>}_{b<}$ with $k$ for different values of $d_i$. As $d_i$ is increased, the Hall wavenumber decreases \textit{i.e.}, the Hall contribution becomes effective for lower $k$ values. Correspondingly, the peak value of the flux rates also shift towards smaller wavenumbers with increasing $d_i$. In addition, the wavenumbers where the fluxes change sign also moves to lower values. 
In accordance with the findings in the previous section, $\Pi^{j>}_{b<}$ fluxes are found to be the strongest irrespective of the value of the $d_i$. The other two fluxes, although weaker than $\Pi^{j>}_{b<}$, shows similar behaviour with the change in Hall scale (not shown here). Fig.~(\ref{jobi-vs-total-di}b) shows the total flux rates to the large scale $\bb$-field for different Hall parameters. As expected, the total flux rates also mimic the nature of the most dominating contribution $\Pi^{j>}_{b<}$.

\section{Discussion and conclusions}
\label{Conclusion}
In this paper, we have systematically investigated the spectral space energy transfer due to the Hall effect in three dimensional HMHD turbulence. In particular, we have studied the role of Hall term on growth of large scale magnetic field \textit{i.e.}, the dynamo effect. Extending the previous studies, where a clear backscatter of magnetic energy was observed, here we have further probed into the said backscatter using spectral fluxes. Unlike the previous studies, here the Hall term has been decomposed into two parts $-d_i \left(\bb\cdot\bnabla\right)\bj$ and $d_i\left(\bj\cdot\bnabla\right)\bb$. The current study clearly shows that the first term, representing the $\bj$-$\bb$ interaction, contributes primarily to the dynamo action. Further classification of the transfer rates in scale specific subparts $\Pi^{j>}_{b<}$, $\Pi^{b>}_{b<}$ and $\Pi^{j<}_{b<}$ shows that the large scale $\bb$-field is mainly fed by the small scale $\bj$-field (Fig.~\ref{flux_various}). In particular, near the Hall scale, the maximum energy transfer takes place from small scale $\bj$ to large scale $\bb$-field, whereas the large scale $\bj$-field extracts energy from the large scale $\bb$-field. However, the net effect mimics the nature of $\Pi^{j>}_{b<}$ which is the strongest of the three aforesaid transfers.
All the flux rates change sign at some wavenumber which moves towards the Hall scale with time. This further verifies the dual cascading nature of the Hall effect found in the previous studies \citep{Mininni_2007, Gomez_2010}. Finally, the sum all three contributions clearly shows that the Hall term adds a non-negligible contribution to the small scale dynamo action.
Finally, the global nature of the flux transfer does not change with the change in the ion inertial length $d_i$.

A thorough understanding of different channels of energy transfer in HMHD is extremely useful to quantify the contribution of the Hall term in sub-MHD scale turbulence in space plasmas {\it e.g.} the solar wind, the magnetospheric plasmas etc. Furthermore, the Hall effect can be extremely important in explaining the generation of large scale magnetic fields in several astrophysical bodies through dynamo effect.
In addition,  the Hall effect is also known to introduce non-locality in the energy transfer process. Therefore, it would be interesting to study how the individual parts contribute to the nonlocality. Our study can also be extended to various other plasma systems such as compressible plasmas, electron-MHD, extended-MHD etc. More specifically, the nonlinear terms present in the aforementioned systems can be decomposed in a similar manner to the Hall term and their individual contributions can be studied in the scale-to-scale transfer process using numerical simulations and \textit{in-situ} observational data.

\section*{acknowledgements}
The authors thank Franck Plunian and Nandita Pan for useful discussions. The simulation code is developed by the first author with the help of parallelization schemes given in \citep{Mortensen_2016}. The simulations are performed using the support and resources provided by PARAM Sanganak under the National Supercomputing Mission, Government of India at the Indian Institute of Technology, Kanpur. MKS acknowledges Tarang for cross-referencing the code for flux calculation \citep{Verma_2013}. SB acknowledges DST-INSPIRE faculty research grant (DST/PHY/2017514) and CEFIPRA research grant (6104-01).

\bibliography{main}

\end{document}